# Microstructural control and tuning of thermal conductivity in $La_{0.67}Ca_{0.33}MnO_{3\pm\delta}$

J.A. Turcaud[1,a], K. Morrison[1], A. Berenov[2], N.McN. Alford[2], K.G. Sandeman[1] & L.F. Cohen[1]

[1]Department of Physics, Blackett Laboratory, Imperial College London, SW7 2AZ, London, U.K.
[2]Department of Materials, Imperial College London, SW7 2AZ, London, U.K.
[a]Author to whom correspondence should be addressed. Electronic mail: jeremy.turcaud08@imperial.ac.uk

Manganites are one of only a small number of material families currently being trialled as room temperature magnetic refrigerants. Here we examine the dependence of the thermal conductivity $\kappa$, of $La_{0.67}Ca_{0.33}MnO_{3\pm\delta}$ as a function of density, grain size and silver impregnation around room temperature. We use a simple effective medium model to extract relevant trends in the data and demonstrate a threefold increase in thermal conductivity by silver impregnation.

The magnetocaloric effect is the adiabatic change in temperature $\Delta T_{ad}$ of a material caused by an applied magnetic field and is largest when a magnetic phase transition is induced. Its potential use in efficient and environmentally friendly solid state cooling was brought to the fore by Brown in 1976 [1] and more recently by Pecharsky and Gschneidner in 1997 [2], whose work on first order phase transitions reignited the research field. There have been significant advances, both in such refrigerant materials and the design of magnetic refrigeration systems over the last 10 years that together bring room temperature magnetic cooling closer to realisation [3,4]. Most of the magnetic refrigerant literature focuses on understanding and enhancing entropy changes [5,6,7,8]. However, the implementation of such materials, particularly in active magnetic regenerators (AMRs) requires knowledge of other physical parameters such as thermal conductivity $\kappa$, that play a crucial role in the control of the efficiency of the cooling cycle [9]. Nielsen and Engelbrecht simulated a two-dimensional AMR and found that optimal values of $\kappa$ depend on the working frequency, varying between 7 and 8 $Wm^{-1}K^{-1}$ and perhaps higher if possible. However, apart from studies reporting $\kappa$ in Gd-Dy, Gd-Si-Ge [10,11], Mn-Fe-P-As [11,12], and La-Fe-Si [11], thermal conductivity has not been systematically investigated. It is worth noting that reported values at room temperature range from: 2 $Wm^{-1}K^{-1}$ for MnAs; 5 $Wm^{-1}K^{-1}$ for $Gd_5Si_2Ge_2$; 9 $Wm^{-1}K^{-1}$ for $La(Fe_{0.88}Si_{0.12})_{13}$, to 10 $Wm^{-1}K^{-1}$ for Gd [11]. A secondary detail is that magnetic and thermal hysteresis, which is highly undesirable because of the loss of efficiency in the cooling cycle, can result from low thermal conductivity and/or ineffective heat transfer during measurement [13]. For all of the above reasons it is therefore desirable to examine the magnitude and tunability of the thermal conductivity of important classes of magnetocaloric materials.

Although the magnetic refrigerants used in most prototypes are metallic [7], oxides and especially manganites such as $La_{1-x}Ca_xMnO_{3\pm\delta}$ are being considered as an interesting alternative [14,15] due to cost, ease of processing and manufacturability, although the isothermal entropy change and $\Delta T_{ad}$ are modest [16,17]. The use of manganites in prototype refrigeration experiments has been recently reported [18]. For such materials, the reported thermal conductivity shows a range of rather low values between 1 $Wm^{-1}K^{-1}$ [19] and 2 $Wm^{-1}K^{-1}$ [20]. Grain size dependence was studied previously in some detail, providing partial explanation of this variability in the literature [21], but the influence of grain size was not separated from that of porosity. The current work builds on ref. 21 and furthermore develops a new methodology to tune $\kappa$ in $La_{0.67}Ca_{0.33}MnO_{3\pm\delta}$ (LCMO) by the introduction of silver. We study the separate influences of density (route 1 – air/manganite microstructure) and grain size (route 2 – dense single phase samples) before considering the introduction into LCMO pellets of a second phase of high-$\kappa$ metallic silver (route 3).

TABLE I. Sintering times, densities, grain sizes, mass percentage of silver and symbols of LCMO samples.

| #[a] | Sinter A[b] (Hrs) | Sinter B[c] (Hrs) | Density (±1%) | Grain size (μm) | Ag Wt% | |
|---|---|---|---|---|---|---|
| 11 | 0.5 | None | 77 | < 1 | -- | ■ |
| 12 | 2 | None | 86 | < 1 | -- | ● |
| 13 | 10 | None | 94 | 2.0±0.1 | -- | ▲ |
| 14 | 20 | None | 96 | 2.6±0.1 | -- | ▼ |
| 15 | 40 | None | 97 | 3.2±0.1 | -- | ★ |
| 21 | 0.5 | 0.5 | 97 | 10.3±0.5 | -- | □ |
| 22 | 2 | 2 | 95 | 10.8±1.0 | -- | ○ |
| 23 | 10 | 10 | 96 | 11.8±1.0 | -- | △ |
| 24 | 20 | 20 | 97 | 13.4±1.3 | -- | -- |
| 25 | 40 | 40 | 97 | 15.9±1.1 | -- | -- |
| 31 | 2 | None | 80 | < 1 | 10.2 | ✳ |
| 32 | 2 | None | 79 | < 1 | 11.7 | ✚ |

[a] # is the sample number.
[b] 1300 °C sintering time in hours.
[c] 1500 °C sintering time in hours.

Each route for controlling $\kappa$ was developed on a different set of cylindrical LCMO pellets, 8 mm wide and 3 mm thick. These were prepared from polycrystalline powders synthesized by glycine nitrate process, using a 1:1 ratio of glycine:nitrate. XRD showed formation of the single phase perovskite structure (space group: *Pnma*) with the lattice parameters a=5.455(1) Å, b=7.718(2) Å, c=5.473(2) Å. The observed lattice parameters are close to those reported in the literature (a=5.454 Å, b=7.708 Å, c=5.471 Å) [22]. All of the pellets were pressed at 100 MPa for 1 minute. Route 1 pellets were sintered at 1300 °C for different amounts of time to achieve varying density (Table I) and oxygenated in air at 900 °C for 30 hours. The second set was subjected to the same treatment as the pellets from set 1, followed by a re-sintering step at 1500 °C for varying amounts of time to promote grain growth, before a final oxygenation at 900 °C for 30 hours (route 2). The third set, used for silver impregnation (route 3), comprised of 2 pellets from set 1. Silver was introduced into both pellets by placing them into a melt of $AgNO_3$ at 300 °C for a chosen amount of time. This treatment was followed by the annealing at 500 °C in air for 10 hours resulting in the decomposition of silver nitrate into metallic silver. Finally, the pellets were polished in order to remove the metallic surface coating.

Density measurements were carried out using an Archimedes method to obtain open porosities, closed porosities and specific gravities, which were compared with the theoretical XRD density of the $La_{0.67}Ca_{0.33}MnO_3$ structure, 6.03 g.cm$^{-3}$ (Table I). The weighing methods were also used to assess the percentage of silver diffused inside the samples of set 3 (samples 31 and 32). Backscattered electron microscopy with Energy-Dispersive X-ray spectroscopy (EDX) and XRD analysis were performed on the polished interior of the sample 31, to determine that silver had penetrated the interior of the sample.

Image processing on backscattered electron images were performed using the public domain, Java-based program ImageJ originally developed by Wayne Rasband from the Research Services Branch of the U.S. National Institute of Health.

The average grain size of each sample was estimated by a linear intercept method averaging the number of grains crossing 100 μm lines in a 100X magnification optical microscope image taken after thermal etching at 1400 °C for 0.5 hours [23]. To do so, the pellets were thoroughly polished down to 1 μm. It should be noted that there was no correction applied to take into account a tetrakaidecahedral grain morphology [24].

Thermal conductivity measurements were carried out with a Quantum Design Physical Properties Measurement System (PPMS) thermal transport option used in continuous measurement mode in a 2-probe configuration. The temperature sweep rate was set at 1 Kmin$^{-1}$ when heating and 0.5 Kmin$^{-1}$ when cooling. The same system, configuration and heating/cooling rate were employed to measure electrical conductivity using a 17 Hz AC bias driving current.

The saturation magnetisation, magnetic field-induced entropy change and heat capacity are insensitive to the processing route. In figure 1 we show representative magnetization curves for each processing route (although the heat capacity and entropy change are not shown as this is not the main focus of the paper). The insets to figure 1 also show some of the representative optical microscopy images. For set 1, as the sintering time increases the densities of the samples also increases (indicated in the inset to Fig. 2(a)). For set 2 shown in Fig. 1(b), the density is almost constant for all pellets at 96 +/-1 % and grain size increases with increased re-sintering time at 1500 °C (indicated in the inset to Fig. 2(b)). The inset in Fig. 1(c) shows backscattered electron image of the interior of the silver impregnated sample 32. The grey areas consist of LCM perovskite (as measured by EDX) and white areas are Ag-rich phase. Note that the silver distribution (white regions) is inhomogeneous, as a result of the geometry of the porous network. There are areas predominantly made up of the LCMO matrix (black regions).

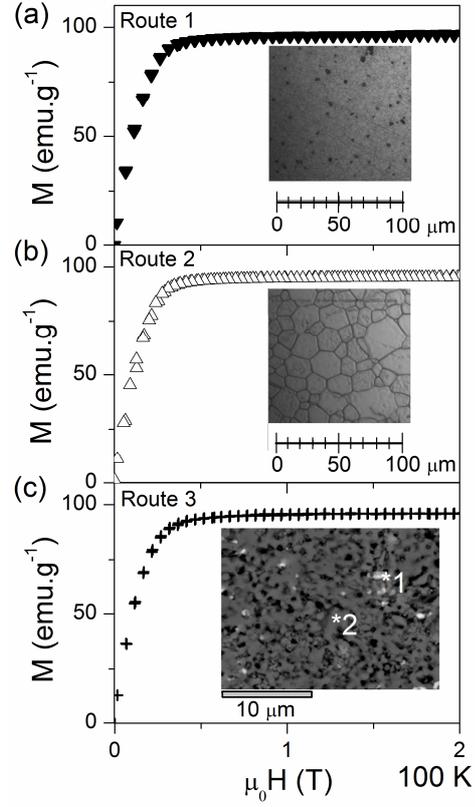

FIG. 1. Saturation magnetisations taken at 100 K of sample 14 synthesized using route 1 in (a), sample 23 synthesized using route 2 in (b), and sample 32 synthesized using route 3 in (c). Inset pictures in (a) and (b) show optical microscope images at 100X magnification of the corresponding samples. The images show grain boundaries (lines) and pores (black spots). Inset picture in (c) is a backscattered electron image of the interior of impregnated sample 32 where the EDX spectra from white areas like (1) are mainly made of silver and the gray areas like (2) are mainly made of LCMO phase. Image processing, using ImageJ software, on this backscattered image reveals a distribution of 4.6 ± 0.7 % of silver, 18.7 ± 3.2 % of pores and 76.7 ± 3.3 % of LCMO.

Fig. 2 shows the temperature dependence of the thermal conductivity of the samples from routes 1 and 2 with (a) increasing density and (b) increasing grain size. The measured $\kappa$ is suppressed at low temperature by the low thermal conductivity of the epoxy paste. Here we focus on trends in $\kappa$ above 100 K where the thermal conductivity of the paste is comparable to that of the samples. It can be seen in Fig. 2 (a) that as the density of the pellets increases, $\kappa$ rises. Using General Effective Media (GEM) theory [25], we model the thermal conductivity of the porous LCMO $\kappa_m$, as a function of the density $D$, using Equation 1 and Equation 2.

$$\kappa_m = \kappa_h (D)^{3/2} \quad \text{for} \quad D > D_c, \quad (1)$$
$$\kappa_m = \kappa_h (1 - \frac{(1-D)}{D_c})^m \quad \text{for} \quad D < D_c, \quad (2)$$

where $\kappa_h$ is the theoretical thermal conductivity of a 100 % dense sample, without any consideration of grain size, and $D_C = 0.838$ corresponds to the critical density where the amount of pores equals the percolation threshold $P_C = 1 - D_C = 0.162$, as established for random close packed media [25]. The exponent *m* in equation (2) is found by fitting to the experimental data to be *m*= 1.89 which belongs to the interval [1.6, 2], defined by the universality class of three dimensional composite materials in which spherical insulating holes are randomly distributed and the total amount of holes is close to the percolation threshold [25]. From the experimental data we can see that $\kappa$ can be tuned by adjusting the density of the sample, e.g. from 1 to 1.5 Wm$^{-1}$K$^{-1}$ at room temperature as shown in the inset in Fig. 2(a).

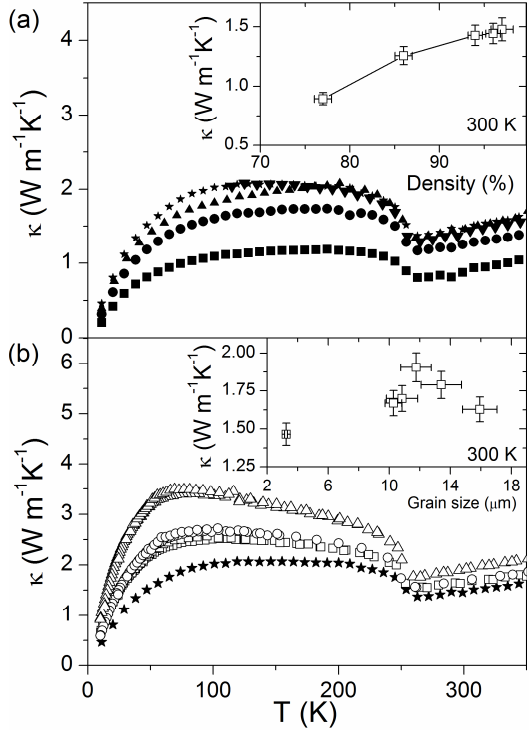

FIG. 2. Thermal conductivity of LCMO pellets with varying density (a) and grain size (b). See Table I for symbol caption. Inset graph in (a) shows that thermal conductivity increases with the density. Inset graph in (b) shows that as the grain size increases, the thermal conductivity tends to rise.

The influence of grain size on the thermal conductivity of LCMO (route 2) is shown in Fig. 2 (b). We see that an increase in the grain size yields an increase in $\kappa$ up to 2 Wm$^{-1}$K$^{-1}$ at room temperature consistent with reports [21], as shown in the inset graph in Fig. 2(b). It also yields an increase in electrical conductivity (not shown here) consistent with reports in other materials [26].

Using density and grain size optimisation in LCMO, the above results demonstrate that we can tune $\kappa$ up to 2 Wm$^{-1}$K$^{-1}$ at room temperature. We have explored whether $\kappa$ can be increased further by inserting a second phase of high thermal conductivity (e.g. metallic silver). Fig. 3 shows the mass percentage of silver that has been taken up by the samples 31 and 32. The top left inset shows open porosities obtained using Archimedes' principle. The top right inset is an XRD spectrum of the cross section, i.e. the inside, of impregnated sample 32 which shows extra peaks from the cubic closed–packed silver phase (space group: *Fm-3m*). These measurements confirm that silver penetrates the samples and it is likely that we are approaching saturation of the total amount of silver that can penetrate the interior of the sample using this route, presumably due to the poor wettability of silver on oxide materials [27].

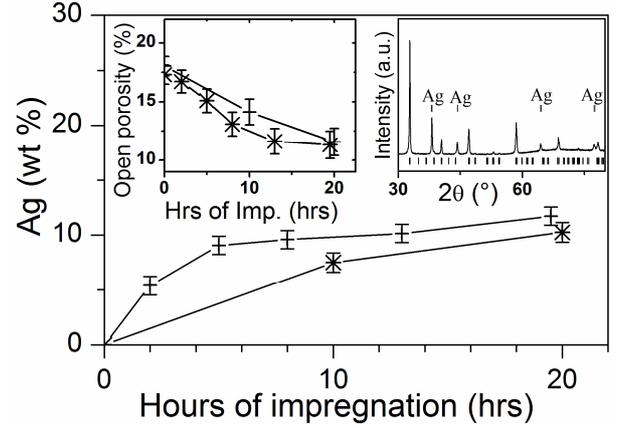

FIG. 3. Mass percentage of silver that has diffused inside LCMO porous samples 31 (asterisks, starting density of 80 %) and 32 (crosses, starting density of 79 %) in the main plot. The upper left inset shows open porosities of samples 31 and 32 as a function of impregnation time. The upper right inset shows an XRD spectrum of sample 32 in which silver is only present as a second phase in addition to the orthorhombic perovskite matrix.

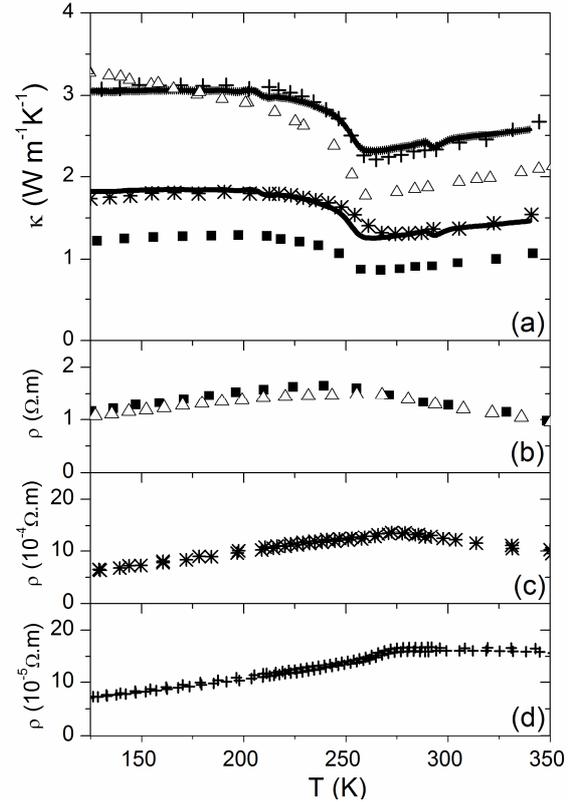

FIG. 4. (a) Thermal conductivity of silver-impregnated sample 31 and 32, (scattered asterisks and crosses respectively), compared with thermal conductivities of the less dense sample, 11 (filled squares) and the grain size-optimized sample 23 (open triangles). The solid lines are the fit to equation 3 for the samples 31 and 32 (see text for details). Electrical resistivity of sample 11 and 23 in (b), of silver-impregnated sample 31 in (c), and sample 32 in (d).

Fig. 4 (a) shows the thermal conductivity of impregnated samples 31 and 32, compared with virgin samples 11 (density 77%) and 23 (optimum grain size and density). We see that silver impregnation raises the room temperature $\kappa$ value to 1.25 Wm$^{-1}$K$^{-1}$ for sample 31 and 2.5 Wm$^{-1}$K$^{-1}$ for sample 32 by a temperature-independent factor. The modest increase of $\kappa$ by a temperature independent factor suggests that silver is not dominating the heat conduction in the impregnated sample as that would have been revealed by a more pronounced increase of $\kappa$ particularly at low temperatures [28]. The resistivity of samples 11, 23, 31 and 32 are shown in Fig. 4 (b), (c) and (d). It can be seen that there is a pronounced decrease of resistivity when silver is added and in addition the characteristic shape of the metal-insulator transition is recovered. Parallel silver conducting paths would reduce the resistivity of course but also short circuit the manganite phase. The fact that we see the opposite suggests that the dominant effect of the silver is to remove the high resistance pathways between the manganite grains. In other words silver has predominantly diffused to the grain boundaries and improved grain to grain connectivity revealing intragrain properties. The distribution of silver can also be explored by employing a GEM analysis of the 3-phase (Air-Ag-LCMO) impregnated samples given by Equation 3:

$$(\kappa_{Composite})^n = (1 - f_{Ag}) \cdot \kappa_{LCMO-porous}{}^n + f_{Ag} \cdot \kappa_{Ag}{}^n, \quad (3)$$

where $\kappa_{Ag}$ is thermal conductivity of pure silver [28], $f_{Ag}$ is the volume proportion of silver in the samples, $\kappa_{LCMO-porous}$ is the simulated starting thermal conductivity of the porous sample before impregnation, using the original density $D$ and Equation 1 or Equation 2 depending on the relative value of $D$ compared to $D_C$ for samples 31 and 32. Finally, $\kappa_{Composite}$ is the measured thermal conductivity of the sample and the logarithmic exponent $n$ determines whether the component parts of the effective medium add in series for which case $n = -1$ or in parallel for $n = +1$ [29]. The GEM model fit is shown in Fig. 4 (a) for the two impregnated samples 31 and 32 yielding $n = 0.08$ and $0.34$ respectively. The positive $n$ exponent indicates that there is a parallel addition of the thermal conductivites of manganite and silver but the non integer value suggests that it is only partial. Taken together these results suggest that silver does not fully percolate throughout the sample volume to provide a completely parallel high thermal conductivity, low resistance path. The fact that the exponent is not unitary suggests that the GEM model used here is likely to be too simplistic to capture the behaviour fully, including the variation in $n$ between samples. EDX mapping of the impregnated sample 32 shows that Ag is certainly distributed inside the pores and possibly at the grain boundaries [30].

In summary, we have studied the influence of density, grain size and the introduction of highly conductive silver on the thermal conductivity of magnetocaloric material, La$_{0.67}$Ca$_{0.33}$MnO$_{3\pm\delta}$. Our study demonstrates a systematic tuning of $\kappa$ from 1 to 2 Wm$^{-1}$K$^{-1}$ at RT by tuning the density and grain size. We have increased $\kappa$ further, up to 2.5 Wm$^{-1}$K$^{-1}$, at RT, using silver impregnation. Such silver addition results in a threefold increase in the thermal conductivity of porous LCMO and represents a method of controlling thermal management in magnetocaloric oxides, or other functional materials that might have a sub-optimal thermal conductivity. It is feasible to apply this technique to improve real magnetic refrigeration devices by diffusing silver into porous preformed plates of manganites. This is an ongoing part of future work. To increase $\kappa$ in these materials still further, other routes to produce a percolating matrix of high conductivity material may be of interest [31].

This work was supported by the UK EPSRC grant no EP/G060940/1. K.G.S acknowledges financial support from The Royal Society.